\documentclass[a4paper,10pt]{amsart}

\usepackage{amssymb,epsfig,amsmath,amsthm}
\usepackage[font=small,format=plain,labelfont=bf,up,textfont=it,up]{caption}
\usepackage{graphicx,subfig,alltt}
\graphicspath{{./Figs/}}
\numberwithin{equation}{section}
\DeclareMathOperator*{\argmin}{arg\,min}

\begin{document}

\title[Policy Iteration in American Option Pricing]{On the Use of Policy Iteration as an Easy Way of Pricing American Options}

\author{C. Reisinger}
\author{J. H. Witte}

\address{Mathematical Institute\\
University of Oxford}
\date{\today}
\email{\texttt{[\,reisinge\,,witte\,\,]\,@\,maths.ox.ac.uk}}
\thanks{The authors acknowledge support from Balliol College, University of Oxford, the
UK Engineering and Physical Sciences Research Council (EPSRC), and the Oxford-Man Institute of Quantitative Finance, University of Oxford.}

\begin{abstract}
In this paper, we demonstrate that policy iteration, introduced in the context
of HJB equations in \cite{Forsyth_Controlled_HJB_PDEs_Finance}, is an extremely simple generic algorithm for solving linear complementarity problems
resulting from the finite difference and finite element approximation
of American options.
We show that, in general, $O(N)$ is an upper and lower bound on the number of iterations needed to solve a discrete LCP of size $N$.
If embedded in a class of standard discretisations with $M$ time steps, the overall complexity of American option pricing is indeed only $O(N (M+N))$, and, therefore, for $M\sim N$, identical to the pricing of European options, which is $O(MN)$.
We also discuss the numerical properties and robustness with respect to model parameters in relation to penalty and projected relaxation methods.\bigskip

\hspace{-.45cm}\textit{Key Words:} American Option, Linear Complementarity Problem, Numerical Solution, Policy Iteration
\end{abstract}

\maketitle

\newtheorem{theorem}{Theorem}[section]
\newtheorem{la}[theorem]{Lemma}
\newtheorem{cor}[theorem]{Corollary}
\newtheorem{remark}[theorem]{Remark}
\newtheorem{prob}[theorem]{Problem}
\newtheorem{definition}[theorem]{Definition}
\newtheorem{alg}[theorem]{Algorithm}
\newtheorem{prop}[theorem]{Proposition}

\section{Introduction}

An American option is a financial instrument that gives its buyer the right, but not
the obligation, to buy (or sell) an asset at an agreed price at any time up to a certain
time $T$. When working in a partial differential equation (PDE) framework,
the option value $V=V(t,S)$, where $t\in [0,T]$ and $S\in\mathbb{R}^+$ denote time and value of the underlying stock, respectively, is usually (e.g.\,cf.\,\cite{OptionPricing,Shreve_ContTimeModels}) the solution of
a linear complementarity problem (LCP)
\begin{align*}
\mathcal{L}V\geq &\ 0,\\
V\geq &\ P\\
\text{and}\quad\mathcal{L}V\cdot (V-P)=&\ 0
\end{align*}
with terminal condition $V(T,S) = P(S)$, where $\mathcal{L}$ is a linear (parabolic) differential operator and $P=P(S)$ denotes the payoff
of the option. 
Furthermore, if a fully implicit or weighted time-stepping scheme is applied to the operator $\mathcal{L}$ in the above LCP, one usually has to
solve a discrete LCP in the form of Problem \ref{discr_LCP_def} at every time
step (cf.\,\cite{OptionPricing,Seydel_ToolsCompFinance,Wilmott_IntrodQuantFinance}). Here, $N\in\mathbb{N}$ denotes the length of the space grid.
A simple example is given at the start of Section \ref{Sec_Numerics}.

\begin{prob}\label{discr_LCP_def}
Let $A\in\mathbb{R}^{N\times N}$ be an M-matrix, and let $b$, $c\in\mathbb{R}^N$ be vectors.
Find $x\in\mathbb{R}^N$ such that
\begin{align*}
Ax\geq &\ b,\\
x\geq &\ c\\
\text{and}\quad (Ax-b)_i\cdot (x-c)_i=&\ 0,\quad 1\leq i\leq N.
\end{align*}
\end{prob}

In this context, for $z\in\mathbb{R}^N$ and $1\leq i\leq N$, $(z)_i$ is used to denote the $i$-th element of the vector $z$.
Additionally, throughout this paper, for $Z\in\mathbb{R}^{N\times N}$ and $1\leq i\leq N$, we will use $(Z)_i$ to denote
the $i$-th row of the matrix $Z$.
The definition of an M-matrix can be found in \cite{FiedlerSpecialMatrices}; in particular, an M-matrix $Z$ is non-singular with $Z^{-1}\geq 0$, i.e.\ every
element of $Z^{-1}$ is non-negative.
Matrices of this form arise naturally from most discretisation schemes for partial differential equations. For the existence and uniqueness
of a solution to Problem \ref{discr_LCP_def}, see e.g. \cite{CryerSystematicOverrelaxation} or \cite{WitteReisinger_PenaltyScheme_DiscreteControlledHJBEquations}.\bigskip

For more general classes of matrices $A$, LCPs like the one in Problem \ref{discr_LCP_def} have been widely studied from the point of view of linear and quadratic programming (cf.\,\cite{CottleDantzig_LP}), but these often make no particular use of the special structure of $A$ 
arising from the discretisation of a differential operator (cf.\,\cite{Ahn_NonsymLCPs_IterMethods,CryerSystematicOverrelaxation}).
This
makes the projected successive over-relaxation method (PSOR) -- introduced in \cite{CryerSystematicOverrelaxation} -- the most widely used approach in practice for finite difference matrices, in spite of its relatively slow convergence.
\bigskip

In this paper, we aim to demonstrate that the method of policy iteration, developed in \cite{Forsyth_Controlled_HJB_PDEs_Finance} for the numerical
solution of HJB equations, yields a powerful and beautifully simple method for the solution of Problem \ref{discr_LCP_def}.
Policy iteration is based on the interpretation of Problem \ref{discr_LCP_def}
as the discrete HJB equation
\begin{equation*}
\min\{Ax-b,x-c\}=0,
\end{equation*}
or, equivalently and component-wise,
\begin{equation*}
\min_{\phi\in\{0,1\}}\{\phi(Ax-b)_i+(1-\phi)(x-c)_i\}=0,
\end{equation*}
where $\phi$ is a control parameter: 0 for exercise, 1 for continuation in  state $i$.
The availability of policy iteration
as a direct algorithm for American option pricing does not seem to be widely known; after completing this research, however, we were made aware of results derived independently in \cite{Bokanowski} for ``Howard's algorithm'' (motivated by Markov decision processes, \cite{HowardDynamicProgramming}), which also includes specific results on American option pricing.
We begin by an analysis in Section \ref{Sec_PolicyIteration} which is similar to \cite{Bokanowski}, for the convergence of the method in
no more than $N$ steps, and we then give specific examples from option pricing which demonstrate that this bound is sharp; numerical results in Section \ref{Sec_Numerics} illustrate this behaviour.
We also benchmark this approach against PSOR (cf.\,\cite{CryerSystematicOverrelaxation}) and a penalty method (cf.\,\cite{ForsythQuadraticConvergence}), showing the 
competitiveness of the method in practice.\bigskip

The PSOR method is proven to converge for most matrices arising in option pricing applications -- specifically, positive definite ones (cf.\,\cite{Ahn_NonsymLCPs_IterMethods}) -- but the number
of necessary iterations grows substantially for decreasing mesh size. Projected multigrid methods with carefully constructed grid transfer operators, as
in \cite{Reisinger_Wittum_Multigrid_2004,Holtz_Kunoth_BSpline_MonotoneMultigridMethods,Graser_Kornhuber_MiltigridObstacleProblems}, give grid independent convergence rates, and practically require a similarly low number of iterations
as the policy iteration presented here (and lower for fine meshes and/or large time steps), but for considerably higher implementation effort.\bigskip

The penalty method combined with a Newton-type algorithm to solve the penalised equation as proposed in \cite{ForsythQuadraticConvergence} requires only the solution of
a (tridiagonal) linear system per iteration (in one dimension), where the number of iterations is usually very small in applications; this is, subject to limitations that we will discuss, inherited by the policy iteration presented here. In higher dimensions, the
linear systems arising in the penalty and policy iterations can be solved efficiently by a standard (i.e.\,non-projected) multigrid method.\bigskip

For completeness, we now briefly discuss the relation to
various other methods -- not, or only loosely based on the solution of the
discrete LCP --
which have been brought forward for pricing American options in a finite 
difference framework.\bigskip

In the financial industry, a standard approach is to apply the early exercise
right `explicitly', which results in a decoupling of the two inequalities by
first computing a continuation value, $\hat{x}$ say, i.e.\ 
the value of holding and hedging the option,
and taking the maximum of that and the exercise value,
\begin{eqnarray*}
A \hat{x} = b, \quad
x = \max(\hat{x},c).
\end{eqnarray*}
Except for fully explicit timestepping schemes, where $A$ is the identity
matrix, $\hat{x}$ is only an approximate solution to the LCP, but it is often remarkably accurate in
practice for sufficiently small timesteps.
It is also interpretable as a Bermudan option
approximation to the American option.\bigskip

One class of methods is based on a more sophisticated splitting of the operators, which approximate the continuous LCP by a sequence of simpler problems
in every time step, rather than solving the discretised LCP directly, e.g. see \cite{Ikonen_Toivanen_OperatorSplitting_AmericanOptions,IkonenToivanen_OperatorSplitting_AmericanOptionsStochVol} and references therein; these methods are typically efficient if the underlying splitting for the
European counterpart is accurate.
However, with these approaches, there is no sense of solving a discrete LCP, which means that the convergence properties of the discretised LCP to the continuous
one (e.g. see \cite{ElliotOckendon,Allegretto_FinElErrorEstimate_AmericanOPtionPricing} and references therein) cannot be utilised, but convergence has to be analysed afresh
for the splitted scheme. To our knowledge, there is no comprehensive convergence analysis for these methods to date.\bigskip

Another class of methods exploits explicit knowledge of the topology of the exercise and continuation regions, either by front-fixing (cf.\,\cite{Nielsen_PenaltyAndFrintFixing_AmericanOptions}), method of
lines coupled with Riccati transformations (cf.\,\cite{Meyer_Hoek_AmericanOptions_MethodOfLines,Chiarella_Kang_Meyer_AmericanOption_StochVol_JumpDiff_MethodOfLines}), or -- on the discrete level -- by partition
of the index set of the linear program as in \cite{Borici_Luthi_FastSolutionLCPS_AmericanPutPricing} and \cite{Brennen_Schwartz_ValuationAmericanPutOption}.
The method proposed here is similar to \cite{Brennen_Schwartz_ValuationAmericanPutOption} in the sense that it is a direct method for the solution of the LCP with a finite number of
iterations, but, differing from \cite{Brennen_Schwartz_ValuationAmericanPutOption}, it is not dependent on any structure of the discrete exercise region.\bigskip

The method proposed in this paper is closely related to the policy iteration in \cite{Forsyth_Controlled_HJB_PDEs_Finance}, adapted to the case of an early exercise option.
Implementationally, it is equivalent to an iterative scheme for variational inequalities in \cite{Hoppe_MultigridVarIneq} and the primal-dual active set strategy used in \cite{HintermullerItoKunisch_PrimalDualAcitveSetStrategy_SemismoothNewtonMethod}, a connection also observed in \cite{Bokanowski}.
As we will
discuss, it can also be seen as the limit for infinite penalty parameter of an adaptation of the standard penalty method in \cite{WitteReisinger_PenaltyScheme_DiscreteControlledHJBEquations} to this case. Interestingly, the formal limit of
the standard method in \cite{ForsythQuadraticConvergence} does not lead to a feasible policy iteration.
The absence of any penalty parameter in the method presented here has some
advantages, since it averts the question about the intensity of the penalisation.
It is conceptually a simpler and arguably more intuitive approach.\bigskip

Finally, we point out that our method extends to most jump models in an obvious way, as it is model independent to the extend that only the M-matrix structure of the discretised equations
matters.

\subsection*{Structure of this Paper} In Section \ref{Sec_PolicyIteration}, we introduce policy iteration in the context of American option
pricing, and we prove
finite convergence in at most $N+1$ steps, where $N\times N$ is the size of the discretisation matrices; we also prove that this bound is sharp and that
monotonicity is an essential requirement. In Section \ref{Sec_Numerics}, we present detailed numerical results. Finally, in Section \ref{PolicyAsPenaltyLimit}, we discuss the relation to semi-smooth Newton iterations for a penalty approximation.

\section{The Method of Policy Iteration}\label{Sec_PolicyIteration}

In this section, we describe how to use policy iteration as an algorithm for solving Problem \ref{discr_LCP_def}. We adapt algorithm and proof from \cite{Forsyth_Controlled_HJB_PDEs_Finance} to the specific situation of the discrete LCP considered here.
We use $I_N$ to denote the identity matrix in $\mathbb{R}^{N\times N}$,
and we consider
\[
(\phi_i (Ax)_i + (1-\phi_i) (x)_i)  - (\phi_i (b)_i + (1-\phi_i) (c)_i)  =
0,\quad 1\leq i\leq N,
\]
where $\phi_i \in \argmin_{\phi\in\{0,1\}}\big\{\phi(A x-b)_i+(1-\phi)(x-c)_i\big\}$, $1\leq i\leq N$, is \emph{an} optimal policy in state $i$ (but not necessarily unique); put differently, the solution $x^*$ to Problem \ref{discr_LCP_def} solves
\[
A^* x^* = b^*,
\]
where $(A^*)_i = \phi_i (A)_i + (1-\phi_i) (I_N)_i$ and
$(b^*)_i = \phi_i (b)_i + (1-\phi_i) (c)_i\,$, $1\leq i\leq N$.

\begin{alg}\label{Policy_Iter_Algorithm}
Let $x^0\in\mathbb{R}^N$. For $x^n$ given, let
$\phi^n \in \mathbb{R}^N$,
$A^n\in\mathbb{R}^{N\times N}$ and $b^n\in\mathbb{R}^N$
be such that, for $1\leq i\leq N$, we have
\begin{align*}
(\phi^n)_i \in&\ \argmin_{\phi\in\{0,1\}}\big\{\phi(A x^n-b)_i+(1-\phi)(x^n-c)_i\big\},\\
(A^n)_i =&\ (\phi^n)_i (A)_i + (1-(\phi^n)_i) (I_N)_i\\
\text{and}\quad (b^n)_i =&\ (\phi^n)_i (b)_i + (1-(\phi^n)_i) (c)_i\,,
\end{align*}
from which follows
\begin{align}
(A^n x^n - b^n)_i = \min_{\phi\in\{0,1\}}\big\{\phi(Ax^n-b)_i+(1-\phi)(x^n-c)_i\big\}.\label{Policy_Iter_Algorithm_Eq1}
\end{align}
Find $x^{n+1}\in\mathbb{R}^N$ such that
\begin{equation}
A^n x^{n+1} = b^n.\label{Policy_Iter_Algorithm_Eq2}
\end{equation}
\end{alg}

In essence, in each step of the iteration in Algorithm \ref{Policy_Iter_Algorithm}, we check pointwise which inequality is violated the most, and solve that one with equality. (The only exception to this is when both inequalities are non-negative, in which case we solve the smaller one with equality; however, as we will see in \eqref{PolicyIterationProp_Theorem_Eq1}, this can only happen when computing $x^1$.)

\subsection{Finite termination and linear complexity}
\label{subsec:finiteterm}

\begin{theorem}\label{PolicyIterationProp_Theorem}
Let $(x^n)^{\infty}_{n=0}$ be the sequence generated by Algorithm \ref{Policy_Iter_Algorithm}. Then
\begin{align*}
x^{n+1}\geq&\ x^n,\quad n\in\mathbb{N},\\
\text{and}\quad x^n=&\ x^*,\quad n\geq\kappa,
\end{align*}
where 
$x^*$ is the solution to Problem \ref{discr_LCP_def} and 
$\kappa\in\mathbb{N}$ is a constant independent of $x^0$.
\end{theorem}
\begin{proof}
It can be found in \cite{FiedlerSpecialMatrices} that, since $A$ is an M-matrix by assumption,  $A^n$ as defined in \eqref{Policy_Iter_Algorithm_Eq1}
is an M-matrix; this means in particular that $A^n$ is non-singular, and, hence, Algorithm \ref{Policy_Iter_Algorithm} is well defined.
Independent of $n\in\mathbb{N}$, there are only $2^N$ different compositions that
can be assumed by $A^n$ and $b^n$, as follows from \eqref{Policy_Iter_Algorithm_Eq1},
consequently there are only $2^N$ different
possible values for $x^n$, $n\in\mathbb{N}$.
From \eqref{Policy_Iter_Algorithm_Eq1} and \eqref{Policy_Iter_Algorithm_Eq2} follows
\begin{equation}
0= A^{n-1}x^n-b^{n-1}\geq \min\{Ax^n-b, x^n-c\}= A^{n}x^n - b^{n},\label{PolicyIterationProp_Theorem_Eq1}
\end{equation}
and therefore
\begin{align*}
A^n(x^{n+1}-x^n)=&\ (A^nx^{n+1}-b^n) - (A^{n}x^n - b^{n})\nonumber\\
\geq&\ (A^nx^{n+1}-b^n) - (A^{n-1}x^n-b^{n-1})=0.
\end{align*}
The M-matrix property of $A_n$\,,
$(A^n)^{-1}\geq 0$, implies that $x^{n+1}\geq x^n$ for $n\in\mathbb{N}$. 
Having established that $(x^n)^{\infty}_{n=1}$ is a monotone sequence in a finite set,
we may set $\kappa=2^{N}$, and we can deduce the existence of a limit $x^*$. It follows directly from \eqref{Policy_Iter_Algorithm_Eq1}
and \eqref{Policy_Iter_Algorithm_Eq2} that $x^*$ solves Problem \ref{discr_LCP_def}.
\end{proof}

\bigskip

In the following corollary, we will equip Algorithm \ref{Policy_Iter_Algorithm} with a tie-breaker: if, in \eqref{Policy_Iter_Algorithm_Eq1}, we have a tie of the form
\begin{equation*}
(Ax^n-b)_i = (x^n-c)_i = \min\big\{(Ax^n-b)_i\,,(x^n-c)_i\big\},
\end{equation*}
then we set $(\phi^n)_i=1$,
\begin{equation}
(A^n)_i=(A)_i\quad\text{and}\quad (b^n)_i=(b)_i,\label{tiebreaker}
\end{equation}
such that $(A^n x^n - b^n)_i = (Ax^n-b)_i\,$.
Now, based on \eqref{tiebreaker}, the following corollary provides a much sharper bound for the maximum number of steps of Algorithm \ref{Policy_Iter_Algorithm}.
In the next section, we will give an example which shows that, in common applications, the obtained order is indeed sharp.

\begin{cor}\label{Cor_LessThanNSteps}
If we use the tie-breaker \eqref{tiebreaker}, then the result of Theorem \ref{PolicyIterationProp_Theorem} holds for $\kappa=N+1$, i.e.\ Algorithm \ref{Policy_Iter_Algorithm} has
finite termination in at most $N+1$ steps.
\end{cor}
\begin{proof}
From \eqref{PolicyIterationProp_Theorem_Eq1}, for $1\leq i\leq N$, $n\in\mathbb{N}$, we have that
\begin{equation}
\min\{(Ax^n-b)_i\,, (x^n-c)_i\}\leq 0.\label{Cor_LessThanNSteps_Eq1}
\end{equation}
From \eqref{tiebreaker} and \eqref{Cor_LessThanNSteps_Eq1} and the monotonicity of $(x^n)^{\infty}_{n=1}$\,, for $1\leq i\leq N$, we may deduce that if $(x^n)_i \geq (c)_i$ and $(A^{n})_i=(A)_i$\,, then
\begin{equation*}
(Ax^{n+1}-b)_i=0\quad\text{and}\quad (x^{n+1})_i\geq (x^{n})_i\geq (c)_i\,,
\end{equation*}
which implies that
$(A^{n+1})_i=(A)_i$\,; iterating the argument, we obtain that
\begin{equation}
 \text{if }(x^n)_i \geq (c)_i\ \wedge\ (A^{n})_i=(A)_i\text{ for }n\in\mathbb{N},\text{ then }(A^m)_i=(A)_i\ \forall\ m\geq n.\label{Cor_LessThanNSteps_Eq2}
\end{equation}
Additionally, we know that
\begin{equation}
(x^n)_i \geq (c)_i\quad\text{for}\ 1\leq i\leq N,\ n\geq2,\label{Cor_LessThanNSteps_Eq2.1}
\end{equation}
since  $(x^2)_i\geq (x^1)_i$ and, if $(x^1)_i<(c)_i$\,, then we have $(Ax^1-b)_i=0$, which implies $(x^2)_i = (c)_i$\,. However, we also note that $A^{n+1}=A^n$ for $n\in\mathbb{N}\cup\{0\}$ implies $x^{n+1}=x^*$.
Altogether, we may say that
\begin{itemize}
\item as long as Algorithm \ref{Policy_Iter_Algorithm} has not yet converged, the linear system solved in \eqref{Policy_Iter_Algorithm_Eq2}
must change in every step,
\item and, based on \eqref{Cor_LessThanNSteps_Eq2} and \eqref{Cor_LessThanNSteps_Eq2.1}, from $x^2$ onwards, every row of the linear system in \eqref{Policy_Iter_Algorithm_Eq2} can change at most once before convergence,
\end{itemize}
which means that Algorithm \ref{Policy_Iter_Algorithm} must converge in no more than $N+2$ steps. But the maximum
number of $N+2$ steps can only be reached if $A^2=I_N$ and $A^*=A$, where $(A^*x^* - b^*) = \min\big\{(Ax^*-b)\,,(x^*-c)\big\}$, which is
a contradiction because we would have had $x^1=x^*$. We may conclude that $\kappa=N+1$.
\end{proof}

We point out that, in practice, we find the performance of Algorithm \ref{Policy_Iter_Algorithm} to be indifferent to
the use of a tie-breaker as introduced in \eqref{tiebreaker}, which is not surprising since -- numerically -- we would expect a tie to be rare.

\subsection{Example: American Put and Other Standard Payoffs}
\label{subsec:put}

It follows from the results in Section \ref{subsec:finiteterm} (see also \cite{Bokanowski}) that policy iteration converges in at most $N+1$ steps, where $N$ is the problem size. We will now show that
in the valuation of the standard American put, the number of iterations is indeed proportional to $N$ normally. Specifically, this behaviour results from the fact that for certain linear segments of the obstacle, as are a common feature of option payoffs, policy iteration moves the discrete exercise boundary only by at most one node per iteration. We demonstrate this on the example of the put payoff, but the result generalises to more general piecewise linear payoffs.

\begin{prop}
\label{prop:ON}
Let the tridiagonal matrix $A$ result from the finite difference discretisation of the Black-Scholes operator,
which is only assumed to be of at least first order consistent.
Let $b_i=c_i= P(S_i) = \max(K-S_i,0)$ the put payoff with strike $K$.
Let $q=\max\{i: S_{i}\le K\}$ and, for the exact solution $x^*$ to Problem \ref{discr_LCP_def},
let $e = \min\{i: (A x^*-b)_{i}=0\} \le q$ (a discrete exercise boundary).
If $x^0=c$, then Algorithm \ref{Policy_Iter_Algorithm} terminates in no less than $q-e$ steps.
\end{prop}
\begin{proof}
We prove by induction that
\begin{equation}
\label{inductionstate}
x_i^n = c_i \text{ for all } i\le q-n,
\end{equation}
where $n$ denotes the $n$-th step of Algorithm \ref{Policy_Iter_Algorithm}.
This is true for $n=0$ by choice of $x^0$.
Now assume that (\ref{inductionstate}) holds for some $n\ge 0$.
Noting that $\mathcal{L} P = r K > 0$ for $S<K$, and that any first order consistent discretisation is exact for linear functions, it follows that $(A x^n - b)_i = r K > 0$ if $x^n_{i-1} = x^n_i = x^n_{i+1} = c_i$ and $i<q$, which is given for $i\le q-n-1$ by (\ref{inductionstate}).
Hence, $\min\{(A x^n - b)_i, (x-c)_i\} =  (x-c)_i = 0$, and,
for $i\le q-n-1=q-(n+1)$, we have $(A^n)_i = (I_N)_i$ and $(b^n)_i = c_i$\,; consequently, $x^{n+1}_i=c_i$ for those $i$, which proves (\ref{inductionstate}) for step $n+1$.
That is to say the discrete free boundary moves by no more than one index in each iteration,
$(A x^n-b)_{e}>0$ for $n<q-e$, and the iteration terminates not before step $q-e$.
\end{proof}

Proposition \ref{prop:ON} demonstrates that, for a put payoff, policy iteration reduces to the following simple method:
nodes are picked systematically as candidates for the discrete exercise boundary, and, if the value function
with this exercise strategy solves the linear complementarity problem, the solution is found; if not,
the next candidate node is tested.
This procedure is similar in spirit to the approaches in \cite{Borici_Luthi_FastSolutionLCPS_AmericanPutPricing} and \cite{Brennen_Schwartz_ValuationAmericanPutOption}.
The advantage of policy iteration, however, is that it does not require any knowledge of the topology of
exercise and continuation regions, but these arise as a by-product.

\begin{remark}\label{Remark_LowerBoundNoOfIterations}
In Proposition \ref{prop:ON}, the number of nodes $q-e$ between the free boundary and the strike will be approximately $N(K-S_e)/S_{max}= O(N)$, if $S_e$ is the exercise boundary and
$S_{max}$ the largest node on a uniform grid with $N$ nodes.
\end{remark}

In the present setting, where the discretisation matrix $A$ can be expected to be tridiagonal, the linear system in each step of Algorithm \ref{Policy_Iter_Algorithm} can be solved
by the Thomas algorithm (cf.\,\cite{Ames_NumMethods_PDEs}) in computational complexity $O(N)$. Since, by Corollary \ref{Cor_LessThanNSteps}, Algorithm \ref{Policy_Iter_Algorithm}
converges in at most $N$ steps, the overall complexity for the solution of Problem \ref{discr_LCP_def} is then at most $O(N^2)$.
If, additionally, Problem \ref{discr_LCP_def} results from a time discretisation with $M$ time steps, the overall complexity
for the American option pricing algorithm will be at most $O(MN^2)$.
Interestingly, it can be shown that, for a relevant class of (monotone) discretisations, the \emph{overall} number of policy iterations, i.e.\ over all time steps, is bounded by $M+N$, such that the overall complexity of the method is $O(N (N+M))$.

\begin{cor}
Consider a monotone, at least first order consistent scheme for the Black-Scholes PDE (e.g.\ implicit Euler, or Crank-Nicolson under a time step constraint, and central differences on a sufficiently fine mesh).
Assume that, in each time step, the solution from the previous time step is used as initial value for the policy iteration.
Then the total number of policy iterations over all time steps $1,\ldots,M$ is exactly $q-e+(M-m)$, where $e$ is the discrete exercise boundary for the American put at time step $M$, and $m$ the
number of time steps in which the discrete free boundary has moved at least one grid point.
\end{cor}
\begin{proof}
That $q-e+(M-m)$ is an upper bound can be shown as in the proof of Proposition 4.6 in \cite{Bokanowski} when accounting for the fact that there may be time steps where the discrete free boundary is stationary. Arguments similar to those in the proof of Proposition \ref{prop:ON} show that it is also a lower bound.
\end{proof}

Although we clearly have implicit schemes in mind, the above results also hold for explicit schemes, for which the free boundary can only move one grid cell per time step by construction.

\subsection{The Role of Monotonicity and the M-Matrix Structure}

The proof of Theorem \ref{PolicyIterationProp_Theorem} crucially requires $A$ to be an M-matrix.
For general matrices $A$, if Algorithm \ref{Policy_Iter_Algorithm} is well defined and converges, it follows from \eqref{Policy_Iter_Algorithm_Eq2} that the limit still solves Problem \ref{discr_LCP_def}.
However, generally, i.e.\ without monotonicity, we obtain no more than subsequence convergence, and in that case we may not find a solution. We will now construct such an example.
\bigskip

It is clear that finite difference matrices resulting from a central difference discretisation of non-degenerate one-dimensional linear elliptic equations are M-matrices, as long as the grid size
is sufficiently small; the discretisation matrices, in every time step, of standard implicit schemes for the corresponding linear parabolic problem, e.g.\ the $\theta$-scheme, are then also M-matrices.
The M-matrix property can thus only fail for large drifts and coarse meshes.
For example, suppose
\begin{equation}
-\mathcal{L}V= \frac{\partial V}{\partial t} + \frac{1}{2}\sigma^2 S^2\frac{\partial^2V}{\partial S^2} + \mu(S)S\frac{\partial V}{\partial S}- rV,\label{BS_EQ_FunnyDrift}
\end{equation}
where $\sigma>0$ and $\mu$ to be determined later.
If, for the sake of this example, we use a four point equidistant space grid consisting of $S_0$\,, $S_1$\,, $S_2$\,, $S_3$\,, where $S_0<S_1<S_2<S_3$\,, and we have known terminal data at $t=T>0$ and Dirichlet boundary conditions $V(S_0\,,t)=V(S_3\,,t)=0$, $t\geq 0$, at $S_0$ and $S_3$\,, then discretising $\mathcal{L}V$ fully implicitly in time (with step size $k$) and with central differences in space (with step size $h$) results in a discretisation matrix
\begin{equation}
\left(
\begin{array}{cc}
1+\frac{k}{h^2}\sigma^2S_1^2+r & - \frac{k}{2h^2}\sigma^2S_1^2 - \frac{k}{2h} \mu(S_1)S_1 \\
-\frac{k}{2h^2}\sigma^2S_2^2 + \frac{k}{2h} \mu(S_2)S_2 & 1+\frac{k}{h^2}\sigma^2S_2^2+r\\
\end{array}
\right).\label{BS_EQ_FunnyDrift_DiscrMat}
\end{equation}
If we pick $\mu$ sufficiently large and such that $\mu(S_1)<0$ and $\mu(S_2)>0$, 
the off-diagonals will be positive and the M-matrix property will fail.
It is easy to check that, for an appropriate combination of parameters $k$, $h$, $\sigma$, $\mu$, $S_1$\,, $S_2$\,, $r$ and payoff vector $c$,
Problem \ref{discr_LCP_def} is given by
\begin{equation*}
A = \left(
\begin{array}{cc}
6 & 8 \\
16 & 8
\end{array}
\right),\ b = \left(
\begin{array}{c}
58\\
64
\end{array}
\right)\quad\text{and}\quad c = \left(
\begin{array}{c}
1\\
5
\end{array}
\right).
\end{equation*}
Since there are only four candidate solutions, $w=(0.6,6.8)^T$, $x=(1,5)^T$, $y=(1,6)^T$ and $z=(3,5)^T$, where
$Aw=b$, $x=c$,
\begin{equation*}
\left(
\begin{array}{cc}
1 & 0 \\
16 & 8
\end{array}
\right)y = \left(
\begin{array}{c}
1\\
64
\end{array}
\right)\quad\text{and}\quad
\left(
\begin{array}{cc}
6 & 8 \\
0 & 1
\end{array}
\right)z = \left(
\begin{array}{c}
58\\
5
\end{array}
\right),
\end{equation*}
one verifies that $z$ is the unique solution
of $f(z) := \min\{Az-b,\,z-c\} = (0,0)^{T}$:
\begin{align*}
f(w)=\left(
\begin{array}{c}
-0.4\\
0
\end{array}
\right), \quad
f(x)=\left(
\begin{array}{c}
-6\\
-1
\end{array}
\right),\quad
f(y)=\left(
\begin{array}{c}
-2\\
0
\end{array}
\right),
\quad\text{and}\quad
f(z)=\left(
\begin{array}{c}
0\\
0
\end{array}
\right).
\end{align*}

Using $w$ as initial guess, Algorithm \ref{Policy_Iter_Algorithm} alternates between $w$ and $y$, never finding the correct value $z$.\bigskip


The chosen `mean-repelling' drift $\mu$ is clearly somewhat unconventional; fortunately, for a constant drift as in the Black-Scholes model or mean-reverting drift $\mu$ as in the pricing of currency options (cf.\,\cite{ValuingForeignCurrencyOptions}), say, or for any (more realistic) discretisation in which $h$ is sufficiently small, the discretisation matrix \eqref{BS_EQ_FunnyDrift_DiscrMat} will still turn out to be an M-matrix, and, thus, our theory applies once again.\bigskip

In more than one dimension, the M-matrix property is typically lost for discretisations of cross-derivatives, independent of the mesh size. In such cases, convergence cannot be guaranteed by the present theory; however, we expect this not to be problematic for fine enough meshes.

\section{Performance for Numerical Examples}\label{Sec_Numerics}

In this section, we compare the numerical performance of Algorithm \ref{Policy_Iter_Algorithm} with two other approaches, namely
PSOR and penalisation (see \cite{CryerSystematicOverrelaxation} and \cite{ForsythQuadraticConvergence}, respectively).\bigskip

We price an American put with strike $K$ in a standard Black-Scholes framework (cf.\,\cite{OptionPricing}), i.e. we have 
\begin{equation}
-\mathcal{L}V= \frac{\partial V}{\partial t} + \frac{1}{2}\sigma^2 S^2\frac{\partial^2V}{\partial S^2} + rS\frac{\partial V}{\partial S}- rV\label{BS_PDE}
\end{equation}
and $P(S) = \max\{K-S,0\}$, and we restrict the allowed asset range to $S\in[0,S_{max}]$, where $S_{max}>K$ is taken to be large.
We discretise \eqref{BS_PDE}
following a standard textbook approach (e.g.\,cf.\,\cite{Seydel_ToolsCompFinance}), using a one-sided difference in time and central differences in space,
working with $M$ time steps and $N$ space steps, and solving backwards in time with a fully implicit scheme. This means that, for every time step, we have to solve a discrete LCP
as given in Problem \ref{discr_LCP_def}, where $A$ represents the finite difference discretisation matrix and is
given by (\ref{DefA}), with elements as in (\ref{firstdiscrmat}) to (\ref{lastdiscrmat}),
$b$ represents the solution vector known from the previous time step,
and $c$ represents the payoff vector. The exact parameters used in our computations can be found in Table \ref{tab:DefaultParam}. Regardless of which algorithm we
use for the solution of Problem \ref{discr_LCP_def}, we only terminate the algorithm if we have found a vector $x\in\mathbb{R}^N$ satisfying
\begin{align*}
\frac{Ax-b}{\|b\|_{\infty}} \geq &\ -tol,\\
\frac{x-c}{\|c\|_{\infty}} \geq &\ -tol\\
\text{and}\quad\Big[\ \frac{|(Ax-b)_i|}{\|b\|_{\infty}}\leq tol\quad \vee\quad \frac{|(x-c)_i|}{\|c\|_{\infty}}\leq &\ tol\ \Big],\quad 1\leq i \leq N,
\end{align*}
for some given tolerance $tol>0$. We implement PSOR exactly as described in \cite{Seydel_ToolsCompFinance},
testing all over-relaxation parameters $w_R\in\{1,1.025,1.05,\ldots,1.875,1.9\}$, and we use a non-linear penalty iteration with
penalty parameter $\rho$ as introduced in \cite{ForsythQuadraticConvergence}.\bigskip

\begin{table}[!ht]
\begin{tabular}{|c|c|c|c|c|c|c|}
\hline
$r$ & $\sigma$ & T & K & $S_{max}$ & tol & $\rho'$\\ \hline
0.05 & 0.4 & 1 & 100 & 600 & 1e-08 & 1e06 \\ \hline
\end{tabular}
\caption{The parameters used for the numerical computations. We use penalty parameter $\rho=\rho'/k$, where $k=T/M$ is the size of the time step, to correct for the scaling implicitly resulting from the time discretisation (cf.\,\cite{Forsyth_InexactArithmetic_DirectControl_PenaltyMethods}).}
\label{tab:DefaultParam}
\end{table}

\subsection{Dependence on Grid Size and Time Step}
\label{subsec:griddep}

The results of our computations for the different methods are summarised in Table \ref{tab:NoOfIteratios}. For different grid sizes, the table
shows the maximum number and the average number of iterations needed to solve the discrete LCP to an accuracy of $tol$
at one time step, and it also includes the overall runtime needed to price the put when solving the discrete LCP for all time steps.
For PSOR, $w_R^*$ denotes the over-relaxation parameter that performed best for that particular grid size.
(The fact that $w_R^*$ changes with the grid size and is not known in advance is certainly a drawback of the PSOR method.)\bigskip

\begin{table}[b]
\begin{tabular}{|l|c|c|c|c|}
\hline
\textit{PSOR} & Max Iterations & $\varnothing$ Iterations & Runtime & $w_R^*$\\
  \hline
$M$, $N=200$ & 3 & 2.17 & 0.23s & 1.050\\
  \hline
$M$, $N=400$ & 5 & 2.67 & 1.07s & 1.075\\
\hline
$M$, $N=800$ & 5 & 3.15 & 4.88s & 1.175\\
\hline
$M=50$, $N=800$ & 28 & 17.00 & 1.61s & 1.650\\
\hline
$M=800$, $N=50$ & 1 & 1.00 & 0.14s & 1.000\\
\hline
\hline
\textit{Penalty Method} & Max Iterations & $\varnothing$ Iterations & Runtime & -\\
  \hline
$M$, $N=200$ & 2 & 1.06 & 0.04s & -\\
  \hline
$M$, $N=400$ & 2 & 1.06 & 0.08s & -\\
\hline
$M$, $N=800$ & 3 & 1.07 & 0.28s & -\\
\hline
$M=50$, $N=800$ & 5 & 1.82 & 0.03s & -\\
\hline
$M=800$, $N=50$ & 2 & 1.00 & 0.07s & -\\
\hline
\hline
\textit{Policy Iteration} & Max Iterations & $\varnothing$ Iterations & Runtime & -\\
  \hline
$M$, $N=200$ & 3 & 1.07 & 0.03s & -\\
  \hline
$M$, $N=400$ & 4 & 1.06 & 0.09s & -\\
\hline
$M$, $N=800$ & 6 & 1.07 & 0.29s & -\\
\hline
$M=50$, $N=800$ & 18 & 2.08 & 0.04s & -\\
\hline
$M=800$, $N=50$ & 2 & 1.00 & 0.07s & -\\
\hline
\end{tabular}
\caption{Our computational results for different numbers of time and space steps, denoted by $M$ and $N$, respectively. We see
that, for all chosen grid sizes, policy iteration and penalisation perform virtually identically and clearly outperform PSOR.}
\label{tab:NoOfIteratios}
\end{table}

We see that, for all considered grid sizes, policy and penalised iteration have very similar
numerical performances and clearly improve over PSOR. Since, for the three considered algorithms, we always solve the discrete LCP at every
time step to the same accuracy $tol$, the numerical results are equivalent to this accuracy, and the computational runtime is the only
criterion that is left to be considered; hence, policy iteration or penalisation should be the methods of choice, with policy iteration
having the (mostly conceptual) advantage of being an exact solver of the LCP rather than an approximation.\bigskip

\begin{table}[b]
\begin{tabular}{|c|c|c|c|c|}
\hline
\textit{Policy Iteration} & $N=100$ & $N=200$ & $N=400$ & $N=800$\\
  \hline
$M=100$ & 2/1.05/0.01s & 4/1.13/0.03s & 7/1.26/0.03s & 14/1.54/0.05s\\
  \hline
$M=200$ & 2/1.03/0.02s & 3/1.07/0.03s & 5/1.13/0.05s & 11/1.27/0.08s\\
\hline
$M=400$ & 2/1.01/0.04s & 2/1.03/0.06s & 4/1.06/0.09s & 8/1.14/0.15s\\
\hline
$M=800$ & 2/1.01/0.06s & 2/1.02/0.11s & 3/1.03/0.16s & 6/1.07/0.29s\\
\hline
\end{tabular}
\caption{Our computational results for different numbers of time and space steps, denoted by $M$ and $N$, respectively.
The three numbers in every cell represent `Max Iterations', `$\varnothing$ Iterations' and `Runtime'.}
\label{tab:NoOfIteratios_MN}
\end{table}
Table \ref{tab:NoOfIteratios_MN} shows that the maximum number of iterations increases linearly with $N$ if $M$ is kept fixed, as predicted by the theory;
it increases with $\sqrt{N}$, if $N=M$, which we will explain in the next subsection.
\bigskip

\begin{figure}[t]
\centering
\includegraphics[width=8cm,height=7cm]{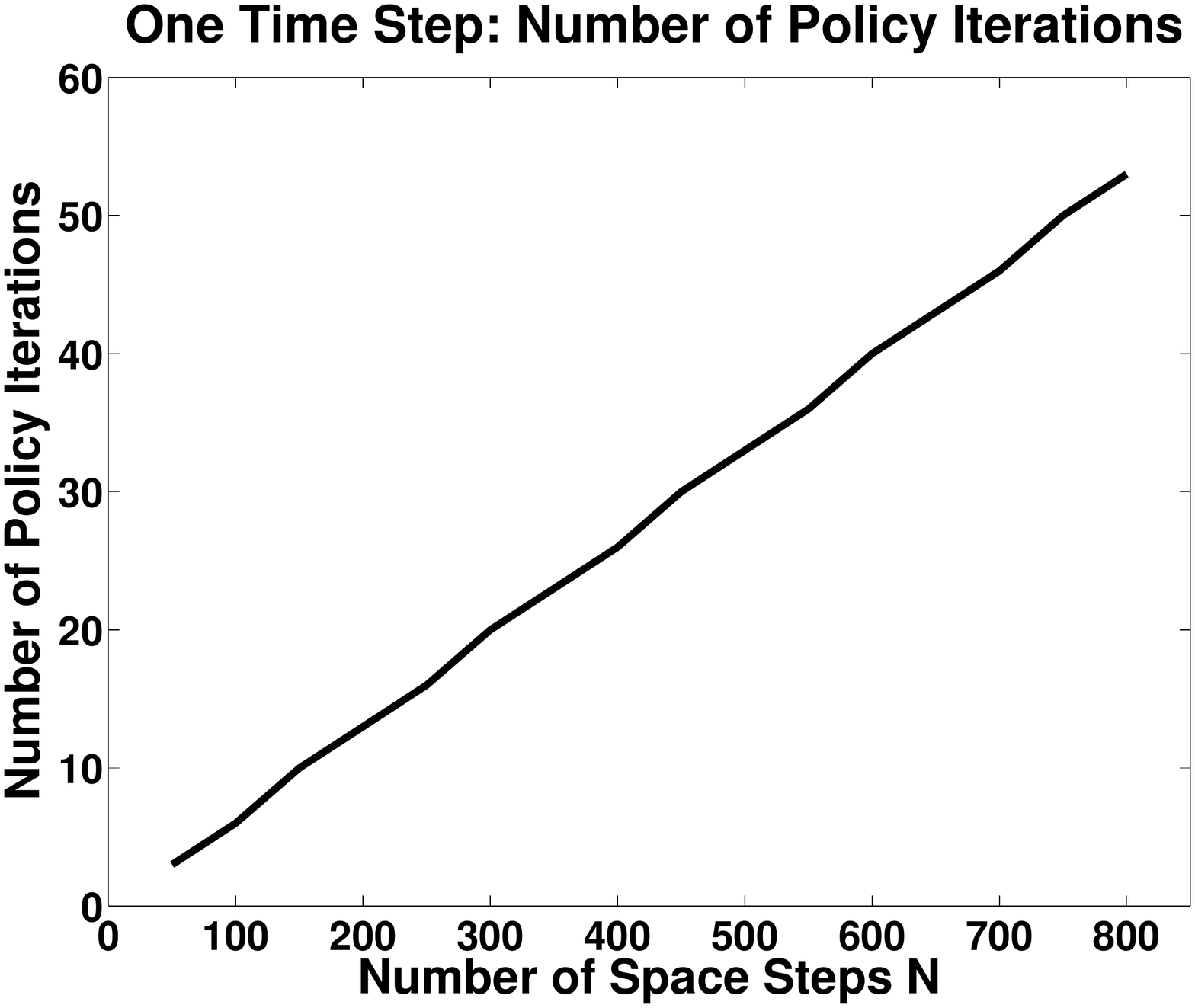}
\caption{We fix $M=1$, solving exactly one discrete LCP, and we consider different grid sizes $N$.
We see that the number of policy iterations required for solving the discrete LCP is clearly linear in the matrix dimension $N$.}
\label{fig:NuOfOtLinear}
\end{figure}

In the light of the results in Section \ref{subsec:put}, we now consider in more detail the number of policy iterations needed to solve a single discrete LCP.
To this end, we set $M=1$ and consider different grid sizes $N$, using $c$ as initial guess.
In Figure \ref{fig:NuOfOtLinear}, we see that the number of policy iterations required is clearly linear in $N$.
\bigskip

Since, in Figure \ref{fig:NuOfOtLinear}, we used one time step only, the starting value of the iteration can be expected
to merely be a rather coarse approximation of the solution. However, the results of Table \ref{tab:NoOfIteratios} suggest
that the average number of
iterations is almost constant if there is a good initial guess, as is the case when using a sufficiently fine time
grid where the active set changes only for a few grid points around the exercise boundary; more precisely,
based on Table \ref{tab:NoOfIteratios}, the observed complexity of American option pricing by penalisation and policy iteration is $O(MN)$, using
$M$ time steps and requiring $O(N)$ for solving a tridiagonal system.\bigskip

In Table \ref{tab:AverageNumOfIt}, we see the average number of policy iterations required when setting $M=N$
and running a fully implicit and a Crank-Nicolson scheme. Again, in both cases, the average number of iterations
is clearly bounded (and very small), here suggesting that the discrete free boundary only moves a small number of grid cells per time step.
The American option pricing algorithm thus shows complexity $O(N^2)$ in practice.\bigskip

As a Crank-Nicolson central difference scheme is hoped to converge with second order
in time and space (at least with a time step selector as in \cite{ForsythQuadraticConvergence} ), $M=N$ is usually the natural choice. Altogether, the observed complexity of an American option pricing algorithm based
on policy iteration (or penalisation) is seen to be the same as for a European pricing code.\bigskip

\begin{table}[t]
\begin{tabular}{|c|c|c|c|c|c|c|c|c|}
\hline
Policy Iteration, $M=N$ & 100 & 200 & 300 & 400 & 500 & 600 & 700 & 800\\ \hline
Fully Implicit, $\varnothing$ Iterations & 1.05 & 1.07 & 1.07 & 1.07 & 1.07 & 1.07 & 1.07 & 1.07 \\ \hline
C.N.\,, $\varnothing$ Iterations & 1.02 & 1.03 & 1.05 & 1.04 & 1.03 & 1.03 & 1.03 & 1.04 \\ \hline
\end{tabular}
\caption{The average number of policy iterations required per time step when setting $M=N$ appears to be constant, for fully implicit as well as Crank-Nicolson (C.N.) time stepping. Hence, in practice, we can price American options in $O(N^2)$, which is the
same complexity as for European options.}
\label{tab:AverageNumOfIt}
\end{table}

\subsection{Dependence on Volatility}

It is known from \cite{BarlesCriticalStockPrice} that, close to expiry, $\sigma^2 (T-t) \ll 1$, the exercise boundary $S_e$ behaves like
\[
\frac{S_e(t)-K}{K} \sim -\sigma \sqrt{-(T-t) \log(T-t)}.
\]

As the number of policy iterations is proportional (cf.\,Remark \ref{Remark_LowerBoundNoOfIterations}) to the displacement of the free boundary from
the initial guess (typically the solution from the previous time step),
we expect that the maximum number of iterations to be attained at the first time step,
where the free boundary moves fastest, and to be proportional to $\sigma$ (cf.\,Table \ref{tab:SigmaSensitivity}).
If $k/h$, and therefore $M/N$, is held constant, one further expects that the maximum number of iterations increases with $\sqrt{N}$, as can be observed in Table \ref{tab:NoOfIteratios_MN}. For large enough $T$, the average number of iterations for all time steps between $t=T$ and $t=0$ will depend weakly on $\sigma$ because of the upper bound $N$ on the total number of iterations over all time steps (cf.\,Proposition \ref{prop:ON} and Table \ref{tab:SigmaSensitivity}).\bigskip

\begin{table}[t]
\begin{tabular}{|c|c|c|c|c|c|c|c|c|}
\hline
$M=1$, $N=200$ & $\sigma=0.2$ & $\sigma=0.4$ & $\sigma=0.8$\\ \hline
Policy Iteration & 6/6 & 13/13 & 23/23\\ \hline
Penalty Method & 4/4 & 6/6 & 7/7\\ \hline
PSOR & 39/39 & 302/302 & 990/990\\ \hline\hline
$M=200$, $N=200$ & $\sigma=0.2$ & $\sigma=0.4$ & $\sigma=0.8$\\ \hline
Policy Iteration & 2/1.03 & 3/1.07 & 6/1.11\\ \hline
Penalty Method & 2/1.03 & 2/1.06 & 2/1.09\\ \hline
PSOR & 2/1.99 & 4/2.18 & 9/2.92\\ \hline
\end{tabular}
\caption{Our computational results for the performance in dependence on $\sigma$. The two numbers in every cell represent `Max Iterations' and `$\varnothing$ Iterations'. For $M=1$, policy iteration has a clear linear dependence on $\sigma$, penalisation is almost unaffected by the change in $\sigma$, and PSOR seems very sensitive to $\sigma$. For $M=N$, all three schemes show only a mild dependence on $\sigma$.}
\label{tab:SigmaSensitivity}
\end{table}

The penalised Newton iteration is least sensitive to changes in $\sigma$, whereas PSOR is affected the most. In the Appendix, we derive the asymptotic convergence rate of PSOR as
$1-c \sigma h^p$, where $p=1$ for $k$ fixed and $p=1/2$ for $k\sim h$.  Consequently, the
asymptotic number of iterations needed to achieve a prescribed accuracy should increase
linearly in $\sigma$, which is broadly consistent with the bottom line of Table \ref{tab:SigmaSensitivity} for the case $M=N=200$. The results for $M=1$ are less clear cut, which may be attributed to $\sigma$ in our example not only affecting the convergence rate, but also
(in a very non-linear way) the distance of the solution to the initial value.

\section{The Relation of Policy Iteration to Penalisation}\label{PolicyAsPenaltyLimit}

Here, we discuss the relation of penalty and policy iterations for American options, and how they can be combined to exploit the  advantages of both.

\subsection{Basic Properties}

The canonical and most common penalty approximation for American options, used e.g.\ in \cite{ForsythQuadraticConvergence} and in the numerical examples of Section \ref{Sec_Numerics}, is obtained by
replacing Problem \ref{discr_LCP_def} by
\begin{equation}
Ax_{\rho} - b - \rho\max\{c-x_{\rho}\,,0\} = 0, \label{Classical_AmericanOp_PenApprox_Eq1}
\end{equation}
for $\rho>0$ large.
A semi-smooth Newton iteration for (\ref{Classical_AmericanOp_PenApprox_Eq1}) can be defined (see \cite{ForsythQuadraticConvergence}) as per Alorithm \ref{Policy_Iter_Algorithm},
but with
\begin{eqnarray}
(\phi^n_\rho)_i &\in& \argmin_{\phi\in\{0,1\}}\big\{(1-\phi)(x_\rho^n-c)_i\big\},\\
(A_\rho^n)_i &=& (A)_i + \rho\, (1- (\phi^n_\rho)_i) (I_N)_i \\
\text{and}\quad(b_\rho^n)_i &=& (b)_i + \rho\,  (1-(\phi^n_\rho)_i) (c)_i\,.
\end{eqnarray}
The number of iterations needed to solve (\ref{Classical_AmericanOp_PenApprox_Eq1}) can be shown to be bounded by $N$, as for
the policy iteration algorithm, and this convergence is also monotone under the same assumptions
(see \cite{ForsythQuadraticConvergence}).
This, and most of the numerical results in Section \ref{subsec:griddep}, suggest similar properties of the two methods, although there is a visible difference in the iteration numbers for large $N$.
Indeed, the iterates generated by the two methods are distinctly different, even for large penalty parameter, which is clear by inspection of the policies  $\phi^n_\rho$ and $\phi^n$.\bigskip

One may be tempted to formally set $\rho=\infty$ in the above equations and define a different
policy iteration by Algorithm \ref{Policy_Iter_Algorithm}, but with the policy $\phi$ replaced by
\begin{eqnarray}
\label{altpolicy}
(\widehat{\phi}^n)_i &\in& \argmin_{\phi\in\{0,1\}}\big\{(1-\phi)(x^n-c)_i\big\}.
\end{eqnarray}
We will now show why this does not result in a convergent algorithm, essentially because the policy does not take into account the PDE constraint.
First, we have to define a tie-breaker for the case $(x^n)_i = (c)_i$\,, when (\ref{altpolicy}) does not uniquely define $(\widehat{\phi}^n)_i$\,.
For the choice $(\widehat{\phi}^n)_i = 0$, the iteration may get `stuck', i.e.\
$(x^m)_i = (c)_i$ for all $m\ge n$, and will not find any solution with $(x^*)_i > (c)_i$\,.
For the choice $(\widehat{\phi}^n)_i = 1$, however, one observes that the solution $x^*$ to Problem \ref{discr_LCP_def} is generally not a fixed point of this policy iteration: if $x^0=x^*$ and $(A x^*-b)_i>0$ for some $i$, then $x_i^0 = x_i^* = c_i$\,, and in the next iteration $(A x^1-b)_i = 0$.\bigskip

We will show in the following section that a slightly modified penalty scheme has policy iteration as its formal limit.\bigskip

It remains to investigate why the number of Newton steps for the
penalised system does not grow in the same way as for policy iteration when the grid is refined,
although the policy algorithm can be viewed as a Newton method for the LCP (see
\cite{Bokanowski}).
Here, it is helpful to take the viewpoint of \cite{ItoKunisch_SemiSmoothNewton_VarIneqFirstKind},
who analyse a semi-smooth Newton method for the underlying continuous variational inequality in
a suitable function space, and show superlinear convergence.
The penalty iteration above can be seen as discretisation of the iteration in
\cite{ItoKunisch_SemiSmoothNewton_VarIneqFirstKind}, and therefore has a well-defined limit
for vanishing grid size.
This is not true for the policy iteration, which is in some sense inherently discrete.
For a discussion of the regularity of the iterates for the unpenalised (continuous) variational inequality,
we refer to \cite{ItoKunisch_SemiSmoothNewton_VarIneqFirstKind}, who also remark on the propagation of discrete active sets, related to the discussion in Section \ref{subsec:put}.

\subsection{Policy Iteration as a Limit of Penalisation}

We briefly show how policy iteration can be seen as limit of a different penalised iteration.



\begin{prob}\label{discr_LCP_Pen_def}
Let $A$, $b$ and $c$ be as in Problem \ref{discr_LCP_Pen_def}. Let $\rho>0$ be large.
Find $x_{\rho}\in\mathbb{R}^N$ such that
\begin{equation}
Ax_{\rho} - b + \rho\min\{Ax_{\rho}-b,x_{\rho}-c\} = 0.\label{discr_LCP_Pen_def_Eq1}
\end{equation}
\end{prob}

\begin{alg}\label{Penalty_Iter_Algorithm}
Let $x^0\in\mathbb{R}^N$. For $x^n_{\rho}$ given, use $A^n$ and $b^n$ as introduced in Algorithm \ref{Policy_Iter_Algorithm}, and
find $x^{n+1}_{\rho}\in\mathbb{R}^N$ such that
\begin{equation}
(A + \rho A^n)x^{n+1}_{\rho} = b + \rho b^n.\label{Penalty_Iter_Algorithm_Eq1}
\end{equation}
\end{alg}

\begin{theorem}\label{Properties_Penalisation}
Problem \ref{discr_LCP_Pen_def} has a unique solution $x_{\rho}$ satisfying
\begin{equation*}
\|x_{\rho}-x^*\|_{\infty}\leq \frac{C}{\rho}
\end{equation*}
for a constant $C>0$ independent of $\rho$, where $x^*$ is the solution to Problem \ref{discr_LCP_def}.
Furthermore, denoting by $(x^n_{\rho})^{\infty}_{n=0}$ the
sequence generated by Algorithm \ref{Penalty_Iter_Algorithm}, it is
\begin{align*}
x^{n+1}_{\rho}\geq&\ x^n_{\rho}\,,\quad n\in\mathbb{N},\\
\text{and}\quad x^n_{\rho}=&\ x^*_{\rho}\,,\quad n\geq\kappa,
\end{align*}
where $\kappa$ is the same
as in Theorem \ref{PolicyIterationProp_Theorem}.
\end{theorem}
\begin{proof}
The results can easily be shown by using the same techniques as in \cite{WitteReisinger_PenaltyScheme_DiscreteControlledHJBEquations}, where more
complex penalty approximations are dealt with.
\end{proof}

Now, we have seen that Problem \ref{discr_LCP_Pen_def} is a viable penalty approximation of Problem \ref{discr_LCP_def} and
that Algorithm \ref{Penalty_Iter_Algorithm} has properties very similar to those of Algorithm \ref{Policy_Iter_Algorithm}.
In addition, equation \eqref{Penalty_Iter_Algorithm_Eq1} bears strong resemblance to equation \eqref{Policy_Iter_Algorithm_Eq1}, especially
if we expect the terms multiplied by $\rho$ to be dominant.

\begin{la}\label{Penalty_Iter_Algorithm_Uniform_Bound}
The sequence $(x^n_{\rho})^{\infty}_{n=0}$ generated by Algorithm \ref{Penalty_Iter_Algorithm} is uniformly bounded in $\rho$, i.e.
\begin{equation*}
\|x^n_{\rho}\|_{\infty}\leq C,\quad n\in\mathbb{N},
\end{equation*}
for a constant $C>0$ independent of $\rho$.
\end{la}
\begin{proof}
For $n\in\mathbb{N}$ and $1\leq i\leq N$, it is
\begin{equation}
(Ax^{n+1}_{\rho}-b)_i=0\label{Penalty_Iter_Algorithm_Uniform_Bound_Eq1}
\end{equation}
or
\begin{equation}
(A x^{n+1}_{\rho} + \rho x^{n+1}_{\rho}-b-\rho c)_i=0.\label{Penalty_Iter_Algorithm_Uniform_Bound_Eq2}
\end{equation}
For equation \eqref{Penalty_Iter_Algorithm_Uniform_Bound_Eq2}, we consider two cases. First, if $(A x^{n+1}_{\rho} - b)_i > (x^{n+1}_{\rho}-c)_i$\,, then we have
\begin{equation}
(x^{n+1}_{\rho} -c)_i \leq 0\leq (A x^{n+1}_{\rho} -b)_i\,.\label{Penalty_Iter_Algorithm_Uniform_Bound_Eq3}
\end{equation}
Second, if $(A x^{n+1}_{\rho} - b)_i \leq (x^{n+1}_{\rho}-c)_i$\,, then it is
\begin{equation}
(A x^{n+1}_{\rho} -b)_i \leq 0\leq (x^{n+1}_{\rho} -c)_i\,.\label{Penalty_Iter_Algorithm_Uniform_Bound_Eq4}
\end{equation}
Now, there is no $\rho$ in equations \eqref{Penalty_Iter_Algorithm_Uniform_Bound_Eq1}, \eqref{Penalty_Iter_Algorithm_Uniform_Bound_Eq3} and \eqref{Penalty_Iter_Algorithm_Uniform_Bound_Eq4}, and
we can deduce the existence of M-matrices $A^*$ and $A^{**}$ such that
$\min\{b,c\} \leq A^* x^{n+1}_{\rho}$ and  $A^{**} x^{n+1}_{\rho}\leq \max\{b,c\}$, where all rows of $A^*$ and $A^{**}$ are either taken
from $A$ or $I_N$; since there are only finitely many compositions that can be assumed by either of $A^*$ and $A^{**}$, we may conclude the proof.
\end{proof}

Based on Lemma \ref{Penalty_Iter_Algorithm_Uniform_Bound}, we can now show that, given identical starting values, one step of policy iteration does in fact
correspond to the limit $\rho\to\infty$ of one step of penalty iteration.

\begin{la}\label{La_x1_difference}
Consider the sequences $(x^n)^{\infty}_{n=0}$ and $(x^n_{\rho})^{\infty}_{n=0}$ generated by Algorithms \ref{Policy_Iter_Algorithm}
and \ref{Penalty_Iter_Algorithm}, respectively. If $x^n = x^n_{\rho}$ for some $n\in\mathbb{N}$, then
\begin{equation*}
\|x^{n+1} - x^{n+1}_{\rho}\|_{\infty}\leq \frac{C}{\rho},
\end{equation*}
where $C>0$ is a constant independent of $n$ and $\rho$.
\end{la}
\begin{proof}
We have
\begin{equation*}
(A+\rho A^n)x^{n+1}_{\rho} = b + \rho b^n\quad\text{and}\quad A^n x^{n+1} = b^n,
\end{equation*}
which implies
\begin{equation*}
\frac{1}{\rho}A x^{n+1}_{\rho} + A^n(x^{n+1}_{\rho} - x^{n+1} )= \frac{1}{\rho}b,
\end{equation*}
and we get the desired result by using the uniform boundedness from Lemma \ref{Penalty_Iter_Algorithm_Uniform_Bound}.
\end{proof}

No matter how large $\rho$ is, Lemma \ref{La_x1_difference} cannot be applied iteratively to compare the whole sequences generated by the two schemes
since it assumes identical starting values. However, we can make the following instructive remark.

\begin{remark}
Loosely speaking, the estimate of Lemma \ref{La_x1_difference} can be interpreted to relate Algorithms \ref{Policy_Iter_Algorithm} and \ref{Penalty_Iter_Algorithm} in
the following sense: if both algorithms use the same starting value $x^0$ and we set $\rho=\infty$, then the generated sequences
are identical. 
\end{remark}

\subsection{A Hybrid Method}

Policy iteration has the conceptual advantage that it solves the discrete LCP exactly, whereas with penalisation there is an additional (small) penalisation error that has to be controlled. 
The Newton iteration for the penalised problem, conversely, has the advantage that it converges faster in practice, and the speed-up can be substantial if $N$ is large. This is an effect of the mechanism by which policy iteration detects the free boundary by a pointwise search.
This suggests a hybrid approach where we use a semi-smooth Newton method to solve the penalised equation and then use this solution as initial value for policy iteration. The following proposition shows that the total number of linear equation solves in this combined algorithm is only by one larger than the Newton steps, with the advantage that the LCP is then solved exactly.

\begin{prop}
\label{penpol}
If $x_\rho$ is the solution of the penalised equation (\ref{Classical_AmericanOp_PenApprox_Eq1}), then for sufficiently large $\rho$ policy iteration with initial value $x_\rho$ converges in a single step.
\end{prop}
\begin{proof}
It suffices to show that the discrete continuation region of the penalty solution $\{i: (x_\rho)_i>c_i\}$ is identical to $\{i: x_i>c_i\}$, i.e.\
\[
x_i > c_i \quad \Rightarrow \quad \exists\ \rho_0 >0 \; \forall\ \rho\ge \rho_0: \;\; (x_\rho)_i > c_i.
\]
But this follows from $\|x-x_\rho\|_\infty \le C/\rho$ (see \cite{ForsythQuadraticConvergence},
\cite{WitteReisinger_PenaltyScheme_DiscreteControlledHJBEquations}).
\end{proof}

\begin{remark}
Asymptotic analysis of the penalisation error of the standard American put and its 
exercise boundary (e.g. see \cite{HowisonReisingerWitte}) suggests that $\rho \sim h^{-2}$ 
is sufficiently large in Proposition \ref{penpol}. This choice is sensible also in terms of overall accuracy of the solution, if the grid convergence is $O(h^2)$ and the penalisation error $O(1/\rho)$.
\end{remark}

\section{Conclusion}

We show that the method of policy iteration, devised in \cite{Forsyth_Controlled_HJB_PDEs_Finance} for the
solution of discretised HJB equations, is a natural
fit to American option pricing. It is extremely simple in structure, and finite convergence in at most $N+1$ steps can easily be proved.
Numerical results show that, in practically relevant situations, it performs identically to a penalty scheme and improves over PSOR.\bigskip

The simplicity advantage of the proposed method is especially noticeable in one dimension, where the algorithm is a small modification of a tridiagonal linear solver; in this
case, the overall complexity for solving the linear complementarity problem is $O(N^2)$ and, if used as part of an instationary American option solver, $O(N(N+M))$ overall.
In higher dimensions, the proposed method is still a direct method in the sense that the basic iteration has finite termination, but the linear systems required by the algorithm might most suitably be solved by an iterative solver.\bigskip

Finally, we discuss how our approach can be interpreted as the formal limit of a Newton-type iteration
applied to a penalised equation, the advantage of the policy iteration being that, in the limit, the penalisation error vanishes.
However, it is also noted that policy iteration is inherently finite-dimensional, and the number of iterations increases for refined meshes, whereas for semi-smooth Newton methods for a zero-order penalty term this is not the case, as there is an underlying continuous iteration which is approximated. 
A consequence of these two points combined is that although two penalty approximations to the same discrete HJB equation may have theoretically identical properties for fixed finite dimensions, 
the infinite dimensional limit is a helpful orientation in that it provides robustness of the method as the grid is refined, a point related to that made in \cite{ItoKunisch_SemiSmoothNewton_VarIneqFirstKind}.

\subsection*{Acknowledgements}
We thank Mike Giles and the two anonymous referees for many helpful comments and advice. We also thank Yves Achdou for bringing reference \cite{Bokanowski} to our attention.

\renewcommand{\bibname}{References}

\bibliography{Paper_HowToSolve_AmericanOption_LCP}
\bibliographystyle{plain}

\appendix

\section{Implementation for Tridiagonal Matrices}\label{PseudoCode}

We briefly describe how Algorithm \ref{Policy_Iter_Algorithm} reduces to a simple variation of
the Thomas algorithm (cf.\,\cite{Ames_NumMethods_PDEs}) if $A$
is a tridiagonal M-matrix; discretisation matrices of tridiagonal structure arise from many problems that are one dimensional
in space (cf.\,\cite{OptionPricing,Seydel_ToolsCompFinance,Wilmott_IntrodQuantFinance}).\bigskip

Consider the situation of Problem \ref{discr_LCP_def}. Suppose matrix $A$ is tridiagonal of the form
\begin{equation}
\label{DefA}
A =
\left(
\begin{array}{cccccc}
\beta_1 & \gamma_1 & & &\\
\alpha_2 & \beta_2 & \gamma_2 & &\\
& \ddots & \ddots & \ddots & \\
& & \alpha_{N-1} & \beta_{N-1} & \gamma_{N-1} \\
& & & \alpha_N & \beta_N & \\
\end{array}
\right),
\end{equation}
and suppose vectors $b$ and $c$ are given by $b=(b_1\,,\ldots,b_N)^T$ and $c=(c_1\,,\ldots,c_N)^T$, respectively.
Moreover, we denote the diagonals of $A$ by $\alpha:=(\alpha_2\,,\ldots,\alpha_N)^T$, $\beta:=(\beta_1\,,\ldots,\beta_N)^T$
and $\gamma:=(\gamma_1\,,\ldots,\gamma_{N-1})^T$. In particular, it is worth pointing out that
$A$ is guaranteed to be an M-matrix
if $\beta> 0$; $\alpha$, $\gamma\leq 0$; all row sums are non-negative; and there is at least
one positive row sum (cf.\,\cite{FiedlerSpecialMatrices}).\bigskip

The following few lines of pseudo-code correspond to an implementation of Algorithm \ref{Policy_Iter_Algorithm} with
starting value $x^0$, solving Problem \ref{discr_LCP_def} to an accuracy of $tol>0$; for notational
convenience, we introduce $\alpha_1:=0$, $\beta_0:=1$, $\gamma_0:=1$, $b_0:=1$ and $\gamma_N:=0$.

\begin{alltt}

1:   SOLVE_LCP\(\,\)(\(x\sp{0},A,b,c,tol\))
2:   \(x\sp{new}\) = \(x\sp{0}\,\)
3:   DO
4:     \(x\sp{old}\) = \(x\sp{new}\)
5:     \(x\sp{new}\) = MODIFIED_THOMAS_ALGORITHM\(\,\)(\(x\sp{old},A,b,c\))
6:   WHILE \(\|x\sp{new}-x\sp{old}\|\sb{\infty}>tol\) 
7:   RETURN \(x\sp{new}\)
8:   END

9:   FUNCTION MODIFIED_THOMAS_ALGORITHM\(\,\)(\(x\sp{old},A,b,c\))
10:  FOR \(i\) = \(1,...\,,N\) DO   
11:    IF (\(x\sp{old}\sb{i-1}\,\alpha\sb{i}+x\sp{old}\sb{i}\,\beta\sb{i}+x\sp{old}\sb{i+1}\,\gamma\sb{i}-b\sb{i}\leq\,x\sp{old}\sb{i}-c\sb{i}\))
12:      \(\lambda\) = \(\alpha\sb{i}/\beta\sb{i-1}\)
13:      \(\beta\sb{i}\) = \(\beta\sb{i}-\lambda\,\gamma\sb{i-1}\)
14:      \(b\sb{i}\) = \(b\sb{i}-\lambda\,b\sb{i-1}\)  
15:    ELSE
16:      \(\beta\sb{i}\) = \(1\,\), \(\gamma\sb{i}\) = \(0\,\), \(b\sb{i}\) = \(c\sb{i}\)
17:    END IF
18:  END FOR
19:  \(x\sp{new}\sb{N}\) = \(b\sb{N}/\beta\sb{N}\)
20:  FOR \(i\) = \(N-1,...\,,1\) DO
21:    \(x\sp{new}\sb{i}\) = (\(b\sb{i}-\gamma\sb{i}\,x\sp{new}\sb{i+1}\))\(/\beta\sb{i}\)
22:  END FOR
23:  RETURN \(x\sp{new}\)

\end{alltt}

As we expect finite termination (cf.\,Theorem \ref{PolicyIterationProp_Theorem}), line 6 presents a meaningful test of convergence.
Alternatively, we
could have checked to what accuracy $x^{new}$ satisfies the LCP in Problem \ref{discr_LCP_def}, which we will do in Section \ref{Sec_Numerics}.\bigskip

The traditional Thomas algorithm is merely the systematic use of Gauss elimination for the solution
of a tridiagonal system of equations, and if lines 11, 15, 16 and 17 were removed from \texttt{MODIFIED\_THOMAS\_ALGORITHM}, the function
would simply be computing $x^{new}$ such that $Ax^{new}=b$. Hence, if we have a European pricing code available, the changes we have to make to account for
American exercise are marginal: we include the function \texttt{SOLVE\_LCP} and slightly modify the existing
Thomas algorithm.

\section{Convergence Rates of PSOR}\label{Appendix_ConvergenceRatePSOR}

Here, we derive an estimate of the convergence rate of PSOR. Although the convergence of PSOR is well documented in the literature, and -- at least for SOR -- precise convergence rates have been derived for certain problems (typically for finite difference discretisations of constant coefficient PDEs), we are not aware of any published estimates for PSOR convergence rates for Black-Scholes-type PDEs.\bigskip

For the Black-Scholes problem, and a fully implicit Euler central difference scheme,
matrix $A$ in  Problem \ref{discr_LCP_def} is tridiagonal as in (\ref{DefA}), with
\begin{eqnarray}
\label{firstdiscrmat}
\alpha_i &=& -k/2 (\sigma^2 i^2 + r i ) \\
\beta_i &=& 1 + k (\sigma^2 i^2 + r ) \\
\text{and}\quad\gamma_i &=& -k/2 (\sigma^2 i^2 - r i ),\quad 1\leq i\leq N,
\label{lastdiscrmat}
\end{eqnarray}
where $k$ denotes the size of the time step, and $A$ is a strictly diagonally dominant M-matrix.
Following \cite{Ahn_NonsymLCPs_IterMethods}, 
PSOR is based on the splitting $A=L+D+U$
into lower triangular, diagonal, and upper triangular matrices, to define an iterative
scheme to solve Problem \ref{discr_LCP_def} by
\[
x^{n+1}-c = \left(x^n-c - \omega D^{-1} (L x^{n+1} + (D+U) x^n-b) \right)^+
\]
for $n\in\mathbb{N}$ and some starting value $x^0$, where $\omega>0$.
It follows from Theorem 4.1 in \cite{Ahn_NonsymLCPs_IterMethods} that the convergence rate
is bounded by the spectral radius $\rho(B)$, where
\[
B = (I-\omega D^{-1} |L|)^{-1} \left| I-\omega D^{-1} (A-L) \right|
=
(I+\omega D^{-1} L)^{-1} \left( I-\omega D^{-1} (A-L) \right)
\]
is the iteration matrix of the SOR method.
Following Young's theorem, see e.g.\ \cite{HackbuschIterativeSolution},
we can determine convergence by analysing the iteration matrix of the Jacobi iteration,
\[
B^{Jac} = I - D^{-1} A.
\]
For $A$ as above, $B^{Jac}$ is tridiagonal with entries
\begin{eqnarray*}
\tilde{\alpha}_i &=& k/2 (\sigma^2 i^2 + r i)/(1 + k (\sigma^2 i^2 + r )), \\
\tilde{\beta}_i &=& 0 \\
\text{and}\quad\tilde{\gamma}_i &=& k/2 (\sigma^2 i^2 - r i )/(1 + k (\sigma^2 i^2 + r )),\quad 1\leq i\leq N,
\end{eqnarray*}
and it follows from Gershgorin's theorem that
\[
\beta: = \rho(B^{Jac}) \le \max_{1\leq i\leq N} (k \sigma^2 i^2)/(1 + k (\sigma^2 i^2 + r )) =
(k \sigma^2 N^2)/(1 + k (\sigma^2 N^2 + r )).
\]
If we let $k N^2\rightarrow\infty$,
\[
\beta \le 1 - \frac{1}{\sigma^2} \frac{1}{k N^2} + O(k^{-2} N^{-4}).
\]
If $k N = S_{max} k/h$ is fixed, $\beta \le 1-c^2/(8 \sigma^2) h + O(h^2)$ for some $c>0$, chosen in this form for notational convenience later.
If $k$ is fixed, $\beta = 1-c^2/(8 \sigma^2) h^2 + O(h^4)$.
We also see that the convergence rate deteriorates with increasing $\sigma$.
Using Young's theorem,
\[
\rho(B) = \left\{
\begin{array}{rl}
1 - \omega + \frac{1}{2} \omega^2 \beta^2 + \omega \beta \sqrt{1-\omega + \omega^2\beta^2/4}, &
0<\omega \le \omega_{opt} \\
\omega-1, & \omega_{opt} \le \omega \le 2,
\end{array}
\right.
\]
where
\[
\omega_{opt} = \frac{2}{1+\sqrt{1-\beta^2}},
\]
and it follows that
\[
\omega_{opt} \le 2 - \frac{c}{\sigma} \sqrt{h} + O(h).
\]
If the leading terms on the right-hand side are chosen for $\omega$,
\[
\rho(B) \le 1- \frac{c}{\sigma} \sqrt{h}.
\]
For $k$ fixed, the number would be $\rho(B) \le 1- \frac{c}{\sigma} h$.
As a consequence, to reduce the error by a factor $\epsilon$, the number of required iterations
is asymptotically
\[
N_{it} \le c \, \sigma \, \frac{\log \epsilon}{\sqrt{h}} 
\]
if $k/h$ is fixed, and $N_{it} \le c \, \sigma \, \frac{\log \epsilon}{h}$ if $k$ is fixed.
A more refined analysis would show that the order of these upper bounds is also sharp.

\end{document}